\documentclass[10pt, conference]{IEEEtran}
\usepackage{times}
\usepackage{graphicx}
\usepackage[keeplastbox]{flushend}
\usepackage{subfig}
\usepackage{amssymb}
\usepackage{amsbsy}
\usepackage{amsmath}
\usepackage{amsthm}
\usepackage{setspace}
\usepackage[applemac]{inputenc}
\usepackage{fancyhdr}
\usepackage{mathtools, cuted}
\usepackage{lipsum, color}
\usepackage{booktabs}
\usepackage{siunitx}

\IEEEoverridecommandlockouts
\usepackage{color,soul}
\usepackage{cite}

\makeatletter
\def\ps@IEEEtitlepagestyle{%
  \def\@oddfoot{\mycopyrightnotice}%
  \def\@evenfoot{}%
}
\def\mycopyrightnotice{%
  {\footnotesize\hfill Accepted for publication in IEEE Communications Letters (DOI 10.1109/LCOMM.2018.2886902). \copyright~2018 IEEE\hfill}%
  \gdef\mycopyrightnotice{}
}
\makeatother

\begin{document}
\title{Impact of LTE's Periodic Interference on Heterogeneous Wi-Fi  Transmissions}
\author{Ilenia Tinnirello, Pierluigi Gallo, Szymon Szott, Katarzyna Kosek-Szott
\thanks{I. Tinnirello and P. Gallo are with University of Palermo, Palermo, Italy.
}
\thanks{S. Szott and K. Kosek-Szott are with AGH University, Krakow, Poland. Their work is supported by contract no. 11.11.230.018. 
}
}

\maketitle

\begin{abstract}
The problem of Wi-Fi and LTE coexistence has been significantly debated in the last years, with the emergence of LTE extensions enabling the utilization of unlicensed spectrum for carrier aggregation. 
Rather than focusing on the problem of resource sharing between the two technologies, in this paper, we study the effects of LTE's structured transmissions on the Wi-Fi random access protocol. We show how the scheduling of periodic LTE transmissions modifies the behavior of 802.11's distributed coordination function (DCF), leading to a degradation of Wi-Fi performance, both in terms of channel utilization efficiency and in terms of channel access fairness.
We also discuss the applicability and limitations of a persistent DCF model in analyzing Wi-Fi performance under periodic LTE interference.  
\end{abstract}

\section{Introduction}
\label{sec:intro}
With most licensed bands being allocated, LTE is now encroaching upon the free spectrum available in unlicensed bands where Wi-Fi is the incumbent technology.
In the struggle for coexistence, LTE's more centralized approach gives it a distinct advantage: without proper measures, Wi-Fi may suffer a 70\%-90\% performance loss \cite{babaei2015impact}.
Therefore, understanding the effects of coexistence is a primary goal for designing next-generation wireless networks. 

An important aspect in the coexistence between Wi-Fi and LTE networks is analyzing how the latter's structured transmissions impact the former's random access approach. LTE defines two mechanisms for the unlicensed bands\cite{zhou2017unlicensed}:
\begin{itemize}
\item \textbf{listen-before-talk} (LBT) -- transmissions are preceded by clear channel assessment and a backoff countdown in case of a busy channel,\item \textbf{carrier sensing adaptive transmission} (CSAT) -- transmissions follow a periodic duty-cycle to allow time division multiplexing between LTE and other technologies (with LTE transmissions scheduled independently of channel status\footnote{The ``carrier sensing'' is done during the CSAT OFF periods to determine channel occupancy by competing technologies and adapt LTE's duty cycle parameters accordingly.}).    
\end{itemize}
Both these approaches can successfully protect Wi-Fi traffic if correctly configured \cite{cano2016unlicensed}. Nonetheless, they can both be classified as periodic interference from the Wi-Fi perspective. CSAT, used in LTE Unlicensed (LTE-U), poses a particular challenge because it does not check whether the channel is idle before starting its transmission. LBT, used in LTE License Assisted Access (LTE-LAA), is regarded as a more friendly and fair alternative than CSAT \cite{Maule2018,Huang2018}. However, it can still be problematic, because it has been demonstrated that the sensing capabilities of the two technologies can be asymmetric and in some cases LTE stations are not able to sense Wi-Fi transmissions \cite{Giuliano2018}. Moreover, most Wi-Fi cards adjust the sensing threshold in case the channel is occupied by a non-Wi-Fi signal, by erroneously considering it as background noise. The result is that the medium can be considered idle while an LTE frame transmission is still ongoing. Therefore, even in case of LTE-LAA, it may happen that Wi-Fi experiences transmission cycles ending with deterministic collisions with LTE frames.  
Such collision events are expected to have a different impact depending on the length of the Wi-Fi transmission in terms of channel occupancy. Heterogeneous frame transmission times can be due to heterogeneous payloads and/or to heterogeneous data rates (employed by the stations as a function of their perceived channel quality).
While the impact of LTE on Wi-Fi has been well studied under homogeneous Wi-Fi frame sizes and data rates \cite{babaei2015impact,Khairy2017,Cano2017,Pang2017,Gao2017}, we look at the phenomena occurring when \emph{heterogeneous} Wi-Fi transmissions experience periodic interference. 
The analysis is done taking LTE-U as the source of periodic interference, but the model and observed phenomena should be applicable to other sources of periodic interference such as LTE-LAA (in the particular case of LTE stations not sensing Wi-Fi ones) or other TDMA-based technologies. 

We begin by analyzing the impact of LTE-U's duty cycle period lengths on the performance of Wi-Fi networks (Section~\ref{s:phenomena}). Then, we discuss the applicability of DCF persistent models to calculate the throughput performance of Wi-Fi stations in case of periodic interference (using \mbox{LTE-U} as a case study) under the assumption of Wi-Fi stations supporting heterogeneous transmission rates (Section~\ref{s:model}). 
We compare Wi-Fi performance using the analytical model and simulations (Section~\ref{s:valid}).
Finally, some conclusions are drawn in Section~\ref{s:conclu}.

\section{Phenomena Caused by Periodic Interference}
\label{s:phenomena}
LTE impact on Wi-Fi has been mainly analyzed by quantifying the channel resources, for example in terms of airtimes, consumed by LTE and scaling the Wi-Fi throughput proportionally to the channel time left to Wi-Fi.
However, there are phenomena, presented in this section, which may further reduce the Wi-Fi efficiency in presence of LTE networks. 

Consider a network with $N$ saturated Wi-Fi stations able to perfectly sense LTE-U transmissions. Let  $X_i$ be the duration of a generic Wi-Fi frame sent by station $i$, $F$ be the duration of the LTE-U ON period (e.g., 10 ms), and $T$ be 
the LTE-U OFF period, which leads to a LTE-U duty cycle of $F/(F+T)$. We also assume that in case of overlapping between Wi-Fi and LTE transmissions, Wi-Fi frames are corrupted as in the case of collisions between two Wi-Fi frames.

There are two macroscopic phenomena related to these periodic LTE transmissions. First, in the case of Wi-Fi frames of different duration, collision probabilities are no more homogeneous in the network, because longer frames are more prone to collisions under a periodic interference source. Indeed, in the last part of the  $T$ interval, whose duration depends on the frame transmission time (the shadowed intervals in Fig.~\ref{fig:model}), the collision probability is deterministically equal to 1 , because any transmission attempts result in a collision with the next scheduled LTE frame.  
Thus, heterogeneous frame transmission times between the stations correspond to heterogeneous intervals
of the channel time that cannot be effectively used for transmissions. Stations with long transmission times perceive a reduced throughput performance because their available channel time is lower, and because of the increment of the average contention windows. This problem cannot be  mitigated even by the addition of uniform transmission opportunities ($TXOP_{limit}$ values) as in the case of dealing with performance anomaly. Within the same TXOP overlapping with an LTE frame, stations employing multiple transmission units with smaller durations can succeed in delivering a number of frames higher than stations employing longer frame durations. Frame successful transmissions can be revealed by using independent acknowledgements per each frame or by sending a block acknowledgement request in the next channel access.

The second phenomenon related to periodic LTE transmissions is that, in the time interval left by LTE, Wi-Fi channel access trials and collision rates are no more uniform in time, as assumed by most DCF models developed from Bianchi's decoupling assumption \cite{Bianchi2000}.  To illustrate this phenomenon, Fig. \ref{fig:model} reports the per-slot collision probability, conditioned by the fact that a channel access trial has been performed at a given time from the previous LTE transmission. The results have been gathered for a scenario with $F=T=1000 \times \sigma$ ($\sigma$ being the slot time length) and $N=30$ stations transmitting frames of 1500~B at a data rate of 54 Mb/s (bottom diagram) or 6 Mb/s (top diagram). From the figure, it is evident that channel access trials are not performed at any time instant within $T$, because collision probability results are clustered only in sub-intervals, with empty spaces in between. Indeed, since the access probability at any slot time is directly related to the average residual backoff at that slot, there are some intervals in which no transmission can start because, deterministically, these intervals are occupied by a Wi-Fi transmission. Only when the $T$ interval allows to perform a number of consecutive channel access trials, frame transmissions can start at any time, as evident in the last part of the $T$ interval for stations employing the 54 Mb/s data rate. 

Table~\ref{t:thr} quantifies the throughput results obtained from ns-3 simulations, with and without the presence of LTE periodic interference, by two coexisting stations employing heterogeneous data rates: 54 Mb/s and 6 Mb/s.  Even in the absence
of LTE interference, DCF does not guarantee throughput fairness for the two classes of stations. 
Indeed, in case of a collision, the contention process of the two stations is not resumed synchronously  (i.e., right after the end of the collision): the slower station awaits the expiration of its ACK timeout, while the faster station has already anticipated this timeout and can immediately proceed to the backoff procedure, once the channel is idle, giving it a competitive advantage. When an LTE-U link with a duty cycle equal to 50\% is active, the throughput division between the two stations strongly depends on  $T$, while the total throughput is generally far from being a mere proportional reduction of the case without interference. 

\begin{table}[t]
\caption{Simulation throughput results (from ns-3) for two contending stations employing heterogeneous data rates with and without LTE interference with a duty cycle of 50\%.}
\begin{center}
\begin{tabular}{cccc}
\toprule
Rate & w/o LTE & \multicolumn{2}{c}{w/ LTE} \\ 
 &  & $T=5$ ms & $T=40$ ms  \\ \midrule
54 Mb/s &  4.6 Mb/s & 4.0 Mb/s & 2.4 Mb/s      \\
6 Mb/s &  4.0 Mb/s & 1.3 Mb/s &  1.9 Mb/s      \\ 
\bottomrule
\end{tabular}
\label{t:thr}
\end{center}
\end{table}

\section{Analysis of Channel Access Performance} \label{s:model}
\begin{figure}[t]
    \centering
    \includegraphics[width=\columnwidth]{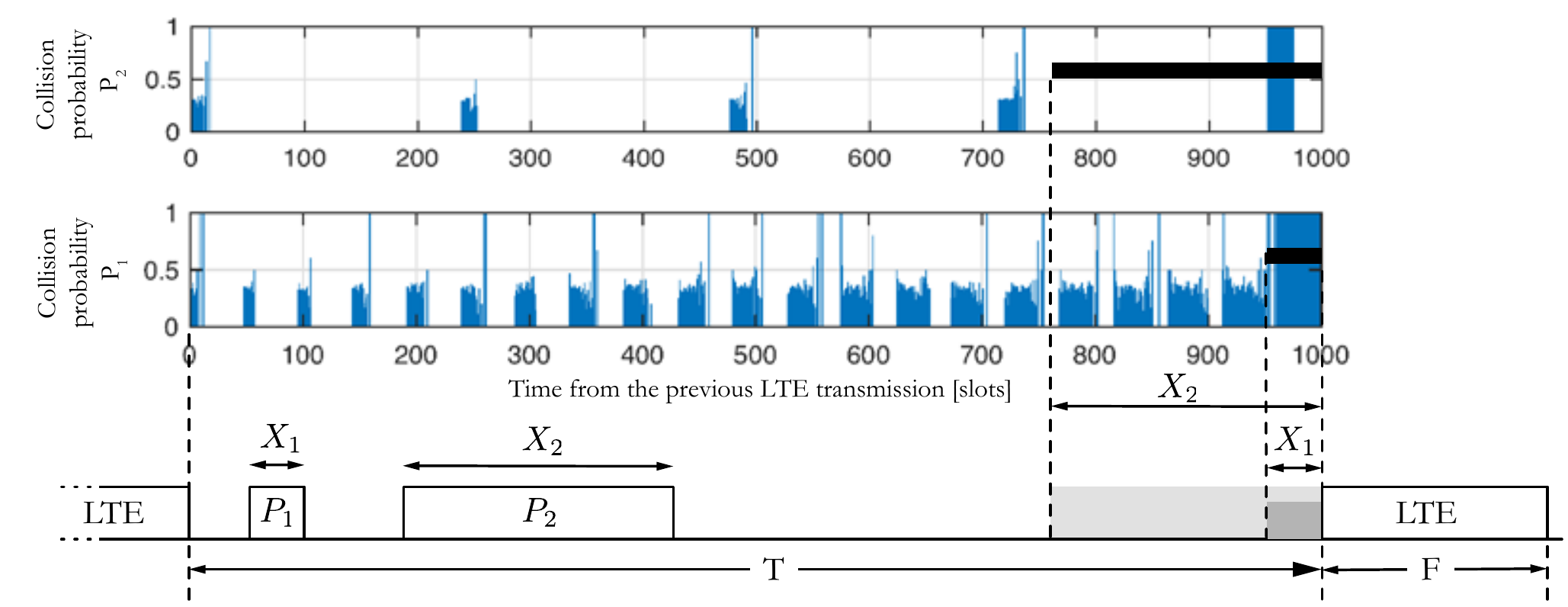}
    \caption{Time sequence of Wi-Fi frames and LTE periodic transmissions and collision rates experienced over time for stations employing heterogeneous transmission times.}
    \label{fig:model}
\end{figure}
Although the phenomena described in the previous section should be addressed by considering a time-varying channel access model, where performance figures, such as the channel access probability $\tau$, depend on the time elapsed from the previous LTE transmission, in this section, we discuss  performance results obtained with a simple adaptation of the DCF Bianchi model and the applicability limits of this approach.

Consider for the sake of simplicity two classes of Wi-Fi stations only, corresponding to two different frame transmissions times $X_1<X_2$. Heterogeneous frame transmission times can be due to heterogeneous payloads and/or to different data rates employed by the stations as a function of their perceived channel quality. Let $n_1$ and $n_2$ be the number of nodes in each class of Wi-Fi stations and let $\tau_1$ and $\tau_2$ be the channel access probability of each class, which we consider time-independent and constant for all the Wi-Fi stations belonging to the same class.  Although we do not model the periodic behaviors of the channel access probability over time, we differentiate the collision rate of each station into two time zones. Thus, for a generic station belonging to class $i$, there is a portion of time (lasting $T-X_i$,) in which there is no risk to interfere with LTE frames and collisions are only due to simultaneous channel accesses attempted by other stations, and a portion of time (lasting $X_i$) in which all  transmissions deterministically end-up with a collision. 
It follows that the average collision probability is non-uniform among the stations, being the portion of time in which it is deterministically equal to 1 given by $X_1$ for one class of stations and $X_2$ for the other one (cf. Fig.~\ref{fig:model}). By considering that each LTE transmission represents a regeneration time for channel access statistics, the collision probability can be given by $p_i=  \frac{T-X_i}{T} \left[1-(1-\tau_i)^{n_i-1}(1-\tau_{-i})^{n_{-i}}\right] + \frac{X_i}{T}=g(\tau_i, \tau_{-i})$, where the index $-i$ refers to the class competing with class $i$ (i.e., $i=1$ and $-i=2$ or vice versa).

For each class of stations, the channel access probability depends on the average contention window ($CW_i$) and the collision probability ($p_i$), i.e., $\tau_i=f(p_i)=\left(1+\frac{1-p_i}{1-p_i^{R+1}}\sum_{j=0}^R p_i^j \frac{CW_j}{2}\right)^{-1}$ as derived in \cite{bianchi2005remarks}, where $CW_i = \min\left(2^{i-1}(CW_{min}+1)-1,CW_{max}\right)$ and $R$ is the max retry limit. 
Channel access and collision probabilities can be derived by solving the non-linear system of equations $\tau_i=f(p_i)$, $p_i=g(\tau_i, \tau_{-i})$ with fixed point iterations. 

For deriving the throughput in b/s, we need to consider the total number of channel access grants $G$ which may result in successful transmissions within each cycle of duration $T+F$. If $E[slot]$ is the average duration of a channel access slot (according to Bianchi's model), then  $G=\frac{T-X_i}{E[slot]}$ for each station of class $i$. Since stations employ heterogeneous data rates and therefore heterogeneous transmission times, the average slot duration can be calculated by considering that the slot lasts $X_1$ with probability $\left[1-(1-\tau_1)^{n_1}\right]  (1-\tau_2)^{n_2}$, $X_2$ with probability $\left[1-(1-\tau_2)^{n_2}\right]$ and $\sigma$ with probability   $(1-\tau_1)^{n_1}  (1-\tau_2)^{n_2}$.

Assuming the two classes of stations $i$ and $-i$ with different transmission rates but same frame payload size $P$, the throughput for class $i$ can be evaluated as:
\begin{equation*}
 S_i = \frac{T-X_i}{E[slot]} \cdot n_i \tau_i (1-\tau_i)^{n_i-1} (1-\tau_{-i})^{n_{-i}} \cdot \frac{P}{T+F}.
\end{equation*}The first factor represents the average number of channel access grants for class $i$ stations which may result in a successful transmission. 
The second factor represents the final probability that the access grant is successful (i.e., only one station accesses the channel).  The last factor takes into account the number of bits transmitted in each channel access and the total duration of the cycle. 

We can also derive the throughput for the case of adopting  uniform $TXOP_{limit}$ values for both station classes (as mentioned in Section~\ref{s:phenomena}).
Assuming that each frame is separately acknowledged, each channel access grant results in a number of transmissions equal to $k_i=\lfloor TXOP_{limit}/X_i \rfloor$ in case of a successful channel access in the time interval  $[0,T-TXOP_{limit}]$, or equal to a number lower than $k_i$ if the channel access is performed in the time interval  $(T-TXOP_{limit},T]$.  Let $k_i^*$ be the average value of the number of successful transmissions performed in the last channel access, that can be approximated by $(k_i-1)/2$ under the simplistic assumption that the channel access is uniform over time at the end of the $T$ interval. 
Although the contention window increasing rate is equalized by the adoption of uniform $TXOP_{limit}$ values, which in turn results in a uniform channel access rate $\tau$ for all the stations, stations employing higher data rates (leading to $k_i>1$) still get an advantage in the final normalized throughput value:  
\begin{align}
\label{eqtxop}
 S_i = \frac{T-TXOP_{limit}}{E[slot]}  \cdot n_i \tau (1-\tau)^{N-1 } \cdot \frac{k_i \cdot X_i }{T+F} + \\ 
 +\frac{n_i \tau (1-\tau)^{N-1 }}{1-(1-\tau)^{N}} \cdot \frac{ k_i^* \cdot X_i}{T+F}, \nonumber
\end{align}
where the last additive term represents the sub-set of frames successfully transmitted in the final channel access grant, which does not result in a collision, before the start of the next LTE interference.

\section{Numerical Results}  \label{s:valid}

\begin{figure*}
\centering
\subfloat[]{\includegraphics[width=0.33\textwidth]{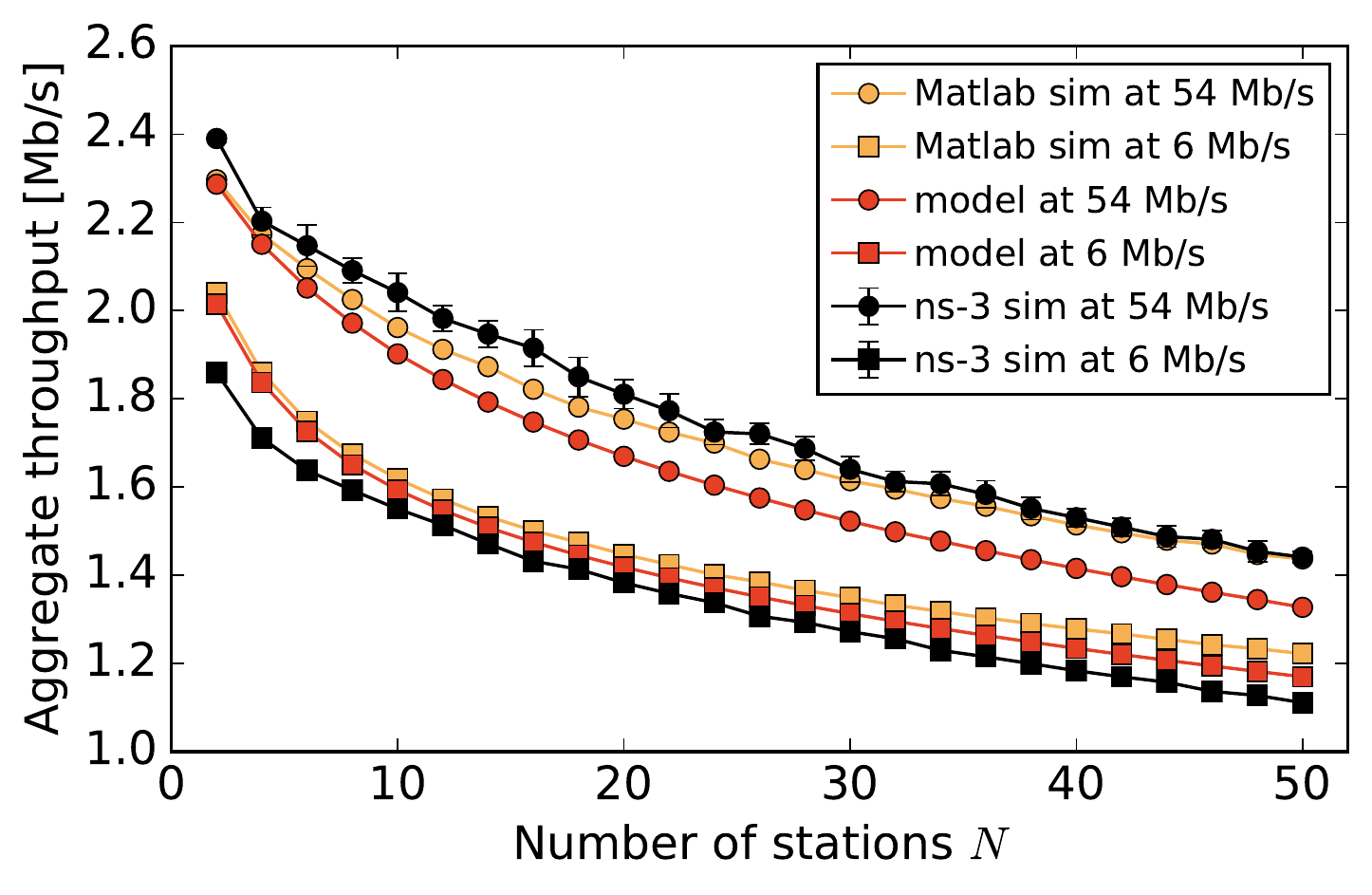}
\label{fig:throughtput}}
\subfloat[]{\includegraphics[width=0.33\textwidth]{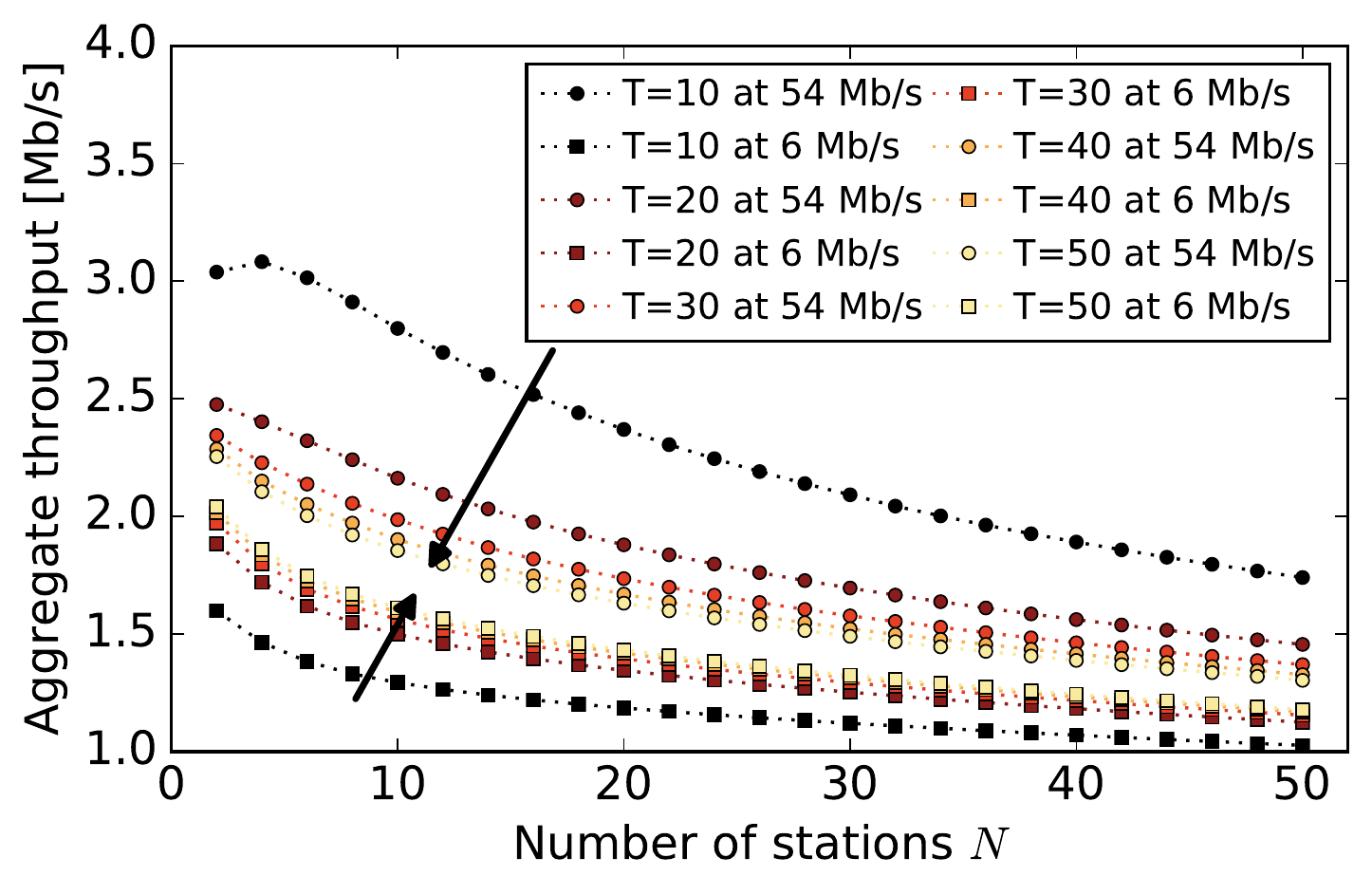}
\label{fig:varyingT}}
\subfloat[]{  \includegraphics[width=0.33\textwidth]{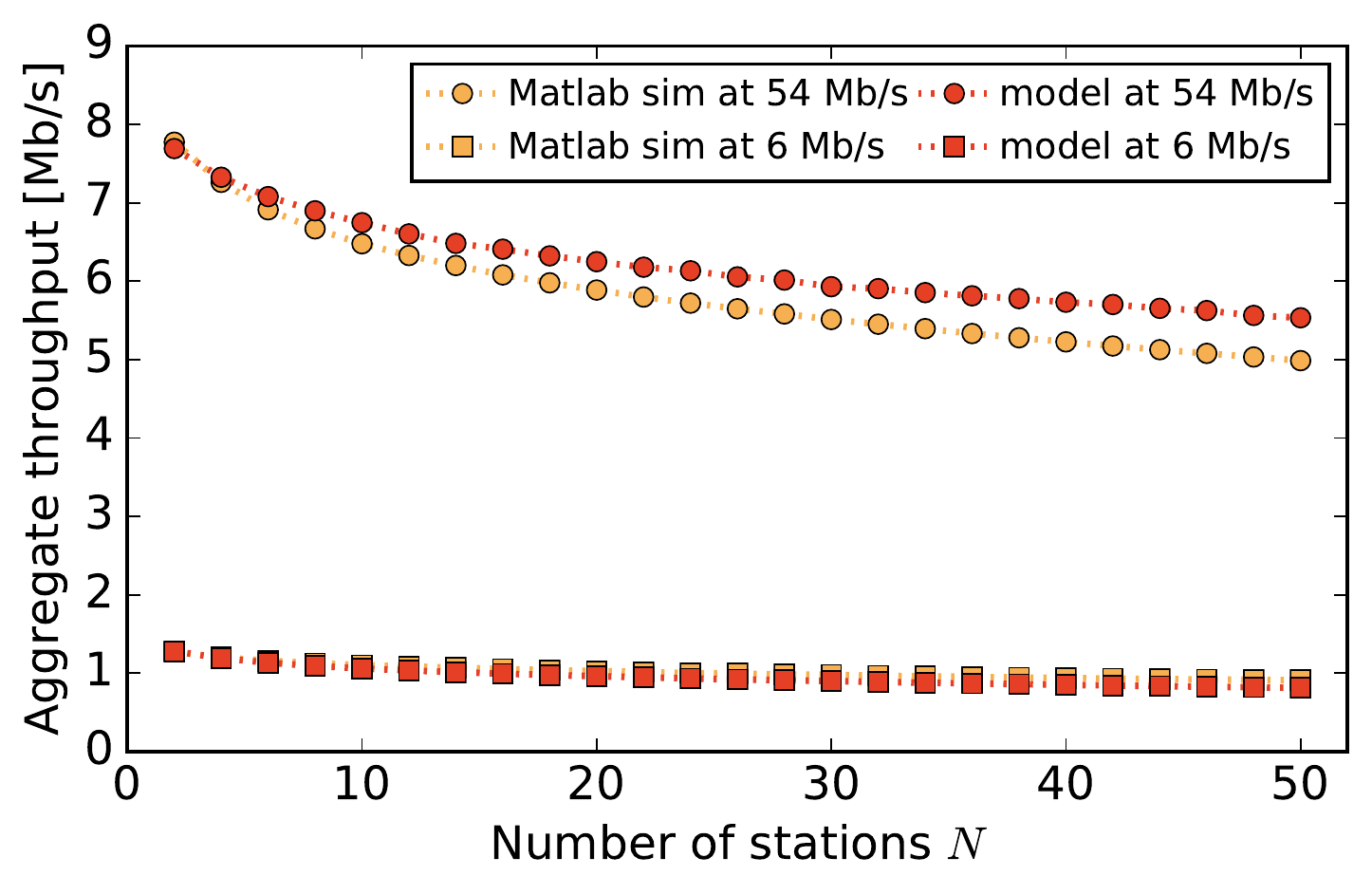}
\label{fig:txop}}
\caption{Performance of  high- and low-rate Wi-Fi stations under periodic LTE interference as a function of the number of stations: (a) aggregate per-class throughput under $F=T=40$ ms for different evaluation methods; (b) throughput dependency on the $F=T$ length; (c) aggregate per-class throughput for uniform setting of $TXOP_{limit}=2.158$ ms. The 95\% confidence intervals are either shown or were too small for graphical presentation.}
\label{fig:comparison}
\vspace{-2ex}
\end{figure*}

In order to demonstrate the phenomena described above, we consider a scenario with a variable total number of Wi-Fi stations $N=2\dots 50$ equally divided into two classes: high-rate (54 Mb/s) and low-rate (6 Mb/s) stations. In this experiment, Wi-Fi stations are subjected to periodic LTE interference. As a performance measure, we analyze the aggregate throughput achieved by each Wi-Fi station class (i.e., high-rate and low-rate).
The system parameters used in our model and in the simulations are indicated in Table \ref{t:parameters}. The results depicted in Fig. \ref{fig:comparison} should be considered as the upper-bound, since all stations always had frames to send.

\begin{table}[]
\centering
\caption{Main Wi-Fi and LTE parameters}
\label{t:parameters}
\begin{tabular}{ll}
\toprule
Wi-Fi Parameter             & Value     \\
\midrule
$CW_{min}$, $CW_{max}$            & 15, 1023        \\
$\sigma$                         & \SI{9}{\micro\second} \\
Retry limit ($R$) 
& 7         \\
Frame size ($P$) & 1500 B\\
$X_1$                       & 0.326~ms (at 54~Mb/s)\\
$X_2$                       & 2.158~ms (at 6~Mb/s)\\
Control rate at 54~Mb/s & 24 Mb/s \\
Control rate at 6~Mb/s & 6 Mb/s\\
$TXOP_{limit}$& 0 (OFF) or 2.158~ms (ON)\\
\midrule
LTE Parameter             & Value     \\
\midrule
LTE-U OFF period  $T$     & 40~ms  or variable   \\
LTE-U ON period   $F$    & 40~ms  or variable  \\
\bottomrule
\end{tabular}
\end{table}

In Fig.~\ref{fig:throughtput} we show the aggregate throughput obtained by high- and low-rate Wi-Fi stations according to
our model, a custom Matlab simulator, and the ns-3.29 simulator (averaged over 10 independent runs). The differences in class throughput confirm the unfairness in channel access and collision probabilities introduced by LTE's periodic interference.

In order to show the effect of varying $T$, we present, in Fig. \ref{fig:varyingT}, throughput results obtained with our model and Matlab simulations. 
The throughput curves are parametrized with values of $T$ between 10 and 50 ms and in all cases $F=T$. The curves for high- and low-rate stations tend to coincide when $T$ increases; in fact, higher values of $T/X_i$ tend to verify the assumption of constant access probability over time, and for $T\to\infty$ the two classes of stations would have the same performance.
Additionally, 
we have calculated the throughput percentage error
${\frac{\mid S_m - S_s\mid}{S_s}\times 100\%}$
between the model and the Matlab simulations for $T\in \{20, 40, 80\}$~ms to validate the presented results.
We found that the model is more accurate for higher values of $T/X_i$ and in all tested cases the observed error was lower than 9\%, which validates the results and confirms the usability of the model.

Finally, in Fig. \ref{fig:txop} we consider setting a uniform $TXOP_{limit}$ value for both classes. Theoretical results and Matlab simulations confirm that  different throughput values are observed for high- and low-rate stations being a result of their different collision probabilities (cf. Eq. \ref{eqtxop}). Thus, setting $TXOP_{limit}$ is not an option to improve fairness.

\section{Conclusions}
\label{s:conclu}
Our analysis of heterogeneous Wi-Fi transmissions coexisting with LTE-U has led us to the following general conclusions regarding Wi-Fi performance under periodic interference:
(i) Wi-Fi frames transmitted just before periodic interference are doomed to collide systematically;
(ii) the throughput degradation caused by periodic interference is more pronounced for longer frames/lower data rates, but it diminishes as the idle time increases;
(iii) periodic interference leads Wi-Fi stations to synchronize their transmissions;
(iv) synchronization occurs in a transitory phase after the periodic interference, where Wi-Fi stations work under impaired random access,
(v) the introduction of uniform $TXOP_{limit}$ values does not improve the observed unfairness.
The observed  Wi-Fi synchronization is not necessarily a problem: secondary synchronization is deliberately forced in multi-hop Wi-Fi networks for mitigating performance impairments due to hidden stations \cite{tinnirello2013supporting}. 
However, in the case of LTE/Wi-Fi coexistence, the synchronization that LTE induces over Wi-Fi is unintentional, and it generates different phenomena.
Therefore, exploiting this synchronization to improve LTE-Wi-Fi coexistence is a possible area of future research.
\bibliographystyle{IEEEtran}

\end{document}